\documentclass[12pt]{revtex4}
  \usepackage{graphics, graphicx}
  \usepackage{color}
  
\begin{document}

 \title{\Large Stephen Hawking:  To Understand the Universe\footnote{Based largely on he authorÕ's contribution to  Physics TodayÕ's March 14 article in which Stephen Hawking is remembered
by his colleagues. \textcolor{blue}{https://goo.gl/dRgf3q)}.}}
 
 \author{James B.~Hartle}

\email{hartle@physics.ucsb.edu}

\affiliation{Department of Physics, University of California,
 Santa Barbara, CA 93106-9530}
 \affiliation{Santa Fe Institute, Santa Fe, NM 87501}

\maketitle

 \large
 
  With the death of Stephen Hawking physicists have lost one of their greatest colleagues and the world has witnessed the conclusion of an inspiring story of triumph over adversity. Personally I have lost a dear friend and matchless collaborator. 

Stephen's major contributions to science are well known and need no review by me; I confine myself to a few personal remarks.

 My association with Stephen began some 46 years  ago during a many month visit I made to Fred Hoyle's Institute of Theoretical Astronomy (as it was known then). In residence were  Brandon Carter, Martin Rees, Paul Davies, and Stephen Hawking --- colleagues with whom I maintained lifelong personal and scientific contacts.  
 In Cambridge I was warmly welcomed by Stephen and Jane.  
 
 From that time on I  always  felt that Stephen and I  were on the same wavelength --- not the same in ability or insight,  of course --- but rather similar in style and  in views of what is important. Ten more joint papers were to follow that visit (see list below). 
 
 For me, the high point of our joint efforts is the paper on the no-boundary wave function of the universe.  
 Stephen wanted to understand the universe in scientific terms.  His deep interest in cosmology runs from his very first papers c. 1965 to his last paper with Thomas Hertog in 2017. 
 
 \begin{figure}[t]
\includegraphics[width=6in]{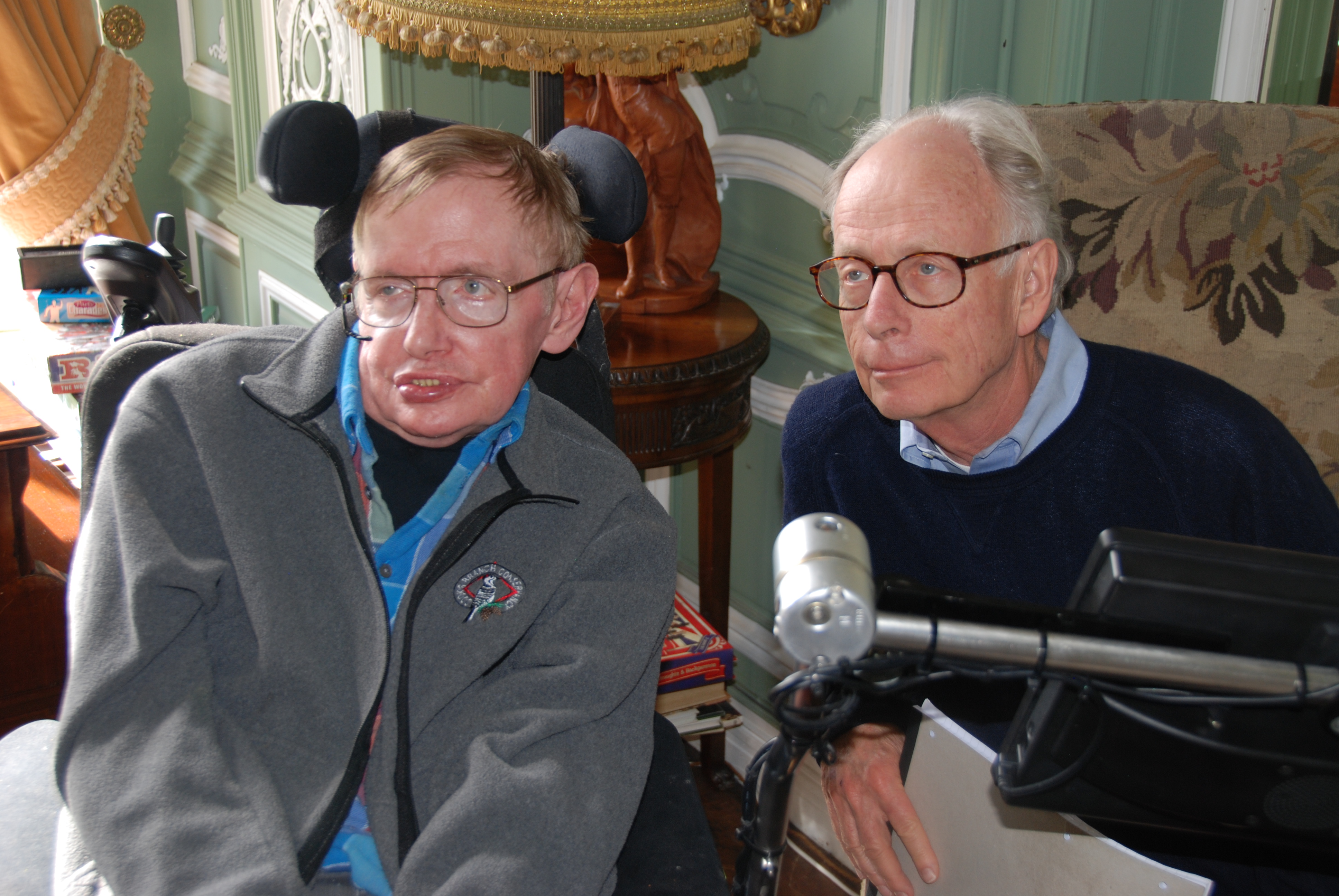}
\caption{Stephen Hawking and James Hartle, Great Brampton House,  Hereford, 2014. \\ Photo courtesy of Don Page}
\label{hawking-hartle}
\end{figure}

  To understand the universe it  is necessary to understand how it began.
 The classical singularity theorems of Stephen and Roger Penrose showed that the universe could not begin with a classical Lorentzian geometry with three space and one time dimension. An earlier joint paper with Stephen  demonstrated the power of Euclidean geometry to help understand the Hawking radiation from black holes. 
  If  the universe couldn't begin classically with a Lorentzian geometry  perhaps it could begin quantum mechanically with a Euclidean geometry. Perhaps it could start with four space dimensions and later make a quantum transition to a Lorentzian spacetime. 
 The result was the no-boundary proposal for the quantum state of the universe. 
 
 I have often thought that the signature of a great problem in physics is that its solution generates more great problems. Certainly that is the case with the no-boundary wave function. The no-boundary wave function of the universe led me to numerous specific calculations, many with Thomas Hertog and Stephen, of what it predicts for our observations of the universe on the largest scales of space and time.   It also motivated a new vision which I  formulated with Murray Gell-Mann  of how  how usual text-book quantum mechanics can be generalized to apply to cosmology.  We called it  decoherent histories quantum theory.  
 
 Working with Stephen was a wonderful experience. He had remarkably clear scientific insight.  He always knew what the right question to ask was. He was able to cut through the clutter that characterizes theoretical physics and focus clearly on the essential points.   Stephen also had the courage to discard  cherished old ideas that are an obstacle to progress like the idea that  black holes are black.  Later when looked at in the right way these seem inevitable. But that was his genius.\footnote{For more about what it was like to work with Stephen see ``Working with Stephen'', arXiv:1711.09071. For some recollections see arXiv:1805.05785}.
 
 How lucky then was my decision as a young assistant professor to take a long leave from Santa Barbara  to work at the Institute of Theoretical Astronomy in Cambridge.  As a consequence, many years later, I consider myself as most  fortunate to have been able to count one of the great scientific figures of the age as a friend and to have been able to work with him on something like an equal basis.  I do not expect to meet his like again. 
 \eject
 
 \appendix
 
 \centerline{\bf Papers with Stephen Hawking}
 
 \begin{itemize}
 
 \item {}  Solutions of the Einstein-Maxwell Equations with Many Black
Holes (with S. W. Hawking), {\sl Comm. in Math Phys}., {\bf 26}, 87-101, 1972.

\item {}  Energy and Angular Momentum Flow into a Black Hole (with S.
W. Hawking), {\sl Comm. Math. Phys.}, {\bf 27}, 283-290, 1972.

\item {}  Path Integral Derivation of Black Hole Radiance (with S. W.
Hawking), {\sl Physical Review D}, { \bf 13}, 2188-2203, 1976.

\item {}  Wave Function of the Universe (with S. W. Hawking), {\sl Physical Review D},  {\bf 28},
2960-2975, 1983. 

\item{} The No-Boundary Measure of the Universe (with S.W. Hawking and T. Hertog),
{\sl Phys. Rev. Lett.},  {\bf 100}, 202301 (2008), arXiv:0711:4630.

\item{} Classical Universes of the No-Boundary Quantum State, (with S.W. Hawking and Thomas Hertog), {\sl Phys. Rev. D},  {\bf  77}, 123537 (2008),\hfil  arXiv:0803:1663.

\item{}  The No-Boundary Measure in the Realm of Eternal Inflation (w. S.W. Hawking and T. Hertog), {\sl Phys. Rev. D}, {\bf 82}, 063510 (2010); \hfil  arXiv:1001:0262.

\item{} Local Observation and Eternal Inflation (w. S.W. Hawking and T. Hertog),  {\sl Phys. Rev. Lett.} {\bf 106}, 141302 (2011);  arXiv:1009.2525.  

\item{}  Vector Fields in Holographic Cosmology, (with S.W.~Hawking and T.~Hertog), {\sl JHEP11}  (2013) 201,  arXiv:1305.719. 

\item{}  Quantum Probabilities for Inflation from Holography, (with S.W.~Hawking, and T.~Hertog),
JCAP, 01(2014) 015,  arXiv:1207.6653.

\item{}  Accelerated Expansion from Negative $\Lambda$, (with S.W.~Hawking, and T.~Hertog), arXiv:1205.3807.  

\end{itemize}

 \end{document}